\documentclass[useAMS]{mn2e}
\usepackage{comment}
\usepackage{url}
\usepackage{color}
\usepackage{subfloat}
\usepackage{graphicx}
\usepackage{caption}
\usepackage{amsmath}
\usepackage{breqn}

\def\hmpc{\;h^{-1}{\rm Mpc}}
\def\hkpc{\;h^{-1}{\rm Mpc}}

\def\hkpc{h^{-1}{\rm kpc}}

\def\lya{Ly$\alpha$}
\def\lyb{Ly$\beta$}
\newcommand{\bL}{\mathbf{L}}
\newcommand{\ee}{\end{equation}}
\newcommand{\bea}{\begin{eqnarray}}                                              
\newcommand{\eea}{\end{eqnarray}}

\newcommand{\bl}{{\pmb \ell}}

\newcommand{\bell}{{\pmb\ell}}

\newcommand{\df}{\delta_{\rm F}}

\def\simlt{\lower.5ex\hbox{$\; \buildrel < \over \sim \;$}}
\def\simgt{\lower.5ex\hbox{$\; \buildrel > \over \sim \;$}}

\def\bi{\bibitem[]{}}

\title[Weak lensing of the Lyman-alpha forest]{
Weak lensing of the Lyman-alpha forest
}

\author[R.~A.~C. Croft {\it et al.}]{Rupert A.~C.~Croft,$^1$\thanks{E-mail: rcroft@cmu.edu} Alessandro Romeo,$^2$,  and R. Benton Metcalf $^{2,3}$\thanks{E-mail: robertbenton.metcalf@unibo.it} \\
$^1$McWilliams Center for Cosmology, Department of Physics, 
Carnegie Mellon University, Pittsburgh, PA 15213, USA\\
$^2$Dipartimento di Fisica e Astronomia, Universita di Bologna,
 via Gabetti 93/2, 4012 Bologna, Italy\\
 $^3$ INAF-Osservatorio Astronomico di Bologna, via Ranzani 1, 40127 Bologna, Italy \\
}

\begin{document}

\pagerange{\pageref{firstpage}--\pageref{lastpage}} \pubyear{2015}

\maketitle

\label{firstpage}

\begin{abstract}
The angular positions of quasars are deflected by the 
gravitational lensing effect of foreground matter.
The \lya\ forest seen in the spectra of these quasars is 
therefore also lensed.
We propose that the signature of weak gravitational lensing of the \lya\ forest
could be measured using similar techniques that have been applied to
the lensed Cosmic Microwave Background, and which have also been proposed for
application to  spectral data from 21cm radio telescopes. 
As with 21cm data, the forest has the advantage of spectral information,
potentially yielding many lensed ``slices'' at different redshifts.
We perform an illustrative idealized test, generating a 
high resolution angular grid of quasars (of order arcminute separation),
and lensing the \lya\ forest spectra at redshifts $z=2-3$
using a foreground density field.
We find that standard quadratic estimators can be used
to reconstruct images of the foreground mass distribution 
at $z\sim1$. There currently
exists a wealth of \lya\ forest data from quasar and galaxy spectral
surveys, with smaller sightline separations expected in the future.
\lya\ forest lensing is sensitive to the foreground mass 
distribution at redshifts intermediate
between  CMB lensing and galaxy shear, and avoids the difficulties
of shape measurement associated with the latter.
With further refinement and application of  mass reconstruction techniques,
weak gravitational lensing of the high redshift \lya\ 
forest may become a useful new cosmological probe.
\end{abstract}


\section{Introduction}
\label{intro}

Gravitational lensing 
has emerged as one of the best ways to probe the
structure of the Universe and to test cosmological models. The distortion
of background images as they are lensed by foreground matter
is sensitive to both the matter contents and the geometry of the Universe
(e.g., Blandford and Narayan 1992, Hoekstra and Jain 2008).
Galaxy images are the most commonly studied cosmological 
sources (see Kilbinger 2015 and references therein),
but the cosmic microwave background (CMB) is also lensed,
and can also be used to reconstruct the foreground lensing
mass distribution (see e.g., the review by Lewis and Challinor
2006). The Planck satellite has recently
enabled all-sky, low fidelity mass maps to be made from CMB lensing (Ade {\it et al.} 2015), 
and the promise of these techniques has prompted interest in  future instruments 
(e.g., Wallis {\it et al.} 2016). Another background source field which has promise
is 21cm radiation from the epoch of reionization (e.g., Combes {\it et al.} 2015),  and from galaxies at lower redshifts (e.g.,  Chang {\it et al.} 2010). 
Techniques used to study CMB lensing have been adapted
and generalized to the three dimensional data expected from future 21cm
surveys (Zahn and  Zaldarriaga 2006 [hereafter ZZ06],  and Metcalf and White 2007 at high redshift, and Pourtsidou and Metcalf 2014 for lower $z$).  In the present paper, we introduce a new background source for weak lensing studies in cosmology, the \lya\ forest.
 
The \lya\ forest of absorption features due to neutral hydrogen can seen
in the spectra of both quasars (Rauch 1992) and galaxies (e.g., Savaglio
{\it et al.} 2002).
We refer to quasars and galaxies as ``backlights'' rather than ``sources''
in what follows, in order to avoid confusion with the ``sources'' 
in gravitational lensing  
(which will be the \lya\ forest here). At the
redshifts ($2 \textless z \textless 6$) where 
the \lya\ transition is in the optical
wavelength range, the forest absorption mostly arises in the 
moderately overdense (of order the cosmic mean) intergalactic medium (IGM)
(Bi 1993, Cen {\it et al.} 1994, Zhang {\it et al.} 1995, Hernquist {\it et al.} 1996).
 This intergalactic medium is a continuous field, and as such
\lya\ forest  spectra can be thought of as a collection of one-dimensional
``intensity maps'' (e.g., Wyithe and Morales 2007) 
of the matter distribution at high redshift. 
Its properties are well studied and it is relatively easy to simulate 
numerically (see e.g., Bolton {\it et al.} 2017). The forest
has been used to test cosmological models, for example
through the influence of the neutrino mass on large-scale structure
(e.g., Palanque-Delabrouille {\it et al.} 2015, Croft {\it et al.} 1999).
 With a high enough angular density 
of quasars, three dimensional statistics can be evaluated by using information
from multiple sightlines, enabling clustering measurements (Slosar {\it et al.} 
2011)
and detection of  baryon
oscillations (Busca {\it et al.} 2013, Slosar {\it et al.} 2013). The collection of 
one-dimensional skewers can also be used to make continuous three dimensional
maps using a variety of interpolation techniques (Pichon {\it et al.} 2001,
Cisewski {\it et al.} 2014), and with even
more numerous star forming galaxies as background spectra these can
be made with angular resolution close to arcminute scales (Lee {\it et al.} 
2014).

The \lya\ forest, being measured from spectra
has  a precisely known source redshift. This fact is also
advantageous for lensing of
the CMB, enabling analyses to be free of uncertainties in the redshift
distribution of sources which affect galaxy weak lensing studies (Hearin
{\it et al.} 2010).
The CMB is also continuous, with statistical properties which make it 
very close to a Gaussian random field. This property has enabled optimal
 estimators of the lensing mass distribution from CMB temperature 
and polarization fields to be constructed (Hu and Okamoto, 2002). 
Observational measurements of CMB lensing were first made by
Das {\it et al.} (2011)
and future instruments are being proposed to take advantage
of the clean nature of the signal and make robust cosmological constraints.

The CMB was the first lensed intensity field to be studied, but lensing of 
21cm radiation has also received much attention. Observational
detections are still in the future, however. 21cm radiation is a 
probe of neutral hydrogen, and at high redshifts, before reionization 
of the IGM, it is a continuous field. At lower redshifts, the residual
HI is mostly found in galaxies and by nature it becomes more Poisson
distributed (Pourtsidou and Metcalf 2014).
The detection of 21cm clustering in cross-correlation
with galaxies by Chang et al (2010) marked the first observational
result in the field of ``intensity mapping''. Many 21cm telescopes
are being built, such as CHIME (Recnik {\it et al.}, 2015) and 
SKA (Combes {\it et al.} 2015, Santos {\it et al.}, 2015), and the possibilities
of 21cm lensing measurements from these are being explored. The effects
of lensing on the statistical properties of the 21cm signal were
studied by Pen (2004) and Cooray (2004).  ZZ06 and Metcalf and White 2007
developed estimators of the lensing mass distribution by analogy with those
developed for CMB lensing. The three dimensional nature of the data (spectral
21cm datacubes) mean that much more information can in principle be 
extracted from 21cm lensing than the CMB, and we can expect that some of
this advantage will be transferred to the \lya\ forest. Many aspects of the 
matter reconstruction techniques for 21cm
(under certain simplifying assumptions) will also apply to the \lya\ forest,
and in this paper we will make use of this to give some illustrative
examples of the potential of \lya\ forest lensing.

Raytracing simulations of 21cm lensing images 
were first made by Hilbert 
{\it et al.} (2007), who added the noise expected from the Metcalf and White 2007 
estimators. Full  simulations of 21cm lensing and
foreground mass reconstruction were made by Romeo (2015),
(see also Romeo {\it et al.}, 2017). Here we will use
the same simulation and reconstruction techniques for this preliminary work
on the \lya\ forest.
In particular, we will make the simplifying assumption that the \lya\ forest
spectra are arranged on a regular grid, so that the FFT based techniques
from 21cm analysis (e.g., ZZ06) can be used without modification. Estimators
of the foreground matter distribution with irregularly spaced
sightlines can be developed (Metcalf {\it et al.}, in prep.), and we plan to 
apply these in future work. 

In this paper we also do not aim to simulate
particular planned surveys or make forecasts. In a companion paper (Metcalf {\it et. al}  2017), however, 
we make quantitive estimates of the expected precision of \lya\ forest lensing measurements for a range of datasets.  Future \lya\ forest surveys
 such as DESI (Aghamousa {\it et al.} 2016)
 will contain close to a million spectra with 
sightlines spaced on average 7-8 arcmins apart. Others, such as the ongoing
CLAMATO survey will cover a small sky area ($\sim 1$ sq. degree) with
higher angular resolution ($\sim 2$ arcmin separation). Here we  carry out
a small simulation test of foreground mass estimation from the forest
to show how 
these techniques can be applied, and leave realistic survey geometries
and  sightline densities to other work (such as Metcalf {\it et al}  2017). We also will only treat widely
separated source and lens fields, and not consider
self-lensing by the forest as was done by Loverde {\it et al.} (2010).

Our outline of the paper is as follows. In Section \ref{lyasource} 
we describe the \lya\ forest as a source field for weak gravitational lensing. 
We explain the relevant lensing geometry and show how it is related to, and different from
the situation for 21cm lensing. In Section \ref{quade} we introduce the lensing
reconstruction quadratic estimator, again taken from 21cm lensing in the
present case. Section \ref{testg} describes our test of gridded data,
outlining how the source and lens fields are generated. Subsection \ref{results}
details the lens reconstruction from the lensed \lya\ forest in our
test and shows results. In Section~\ref{sumconc} we summarize the paper and
conclude with a discussion of what will be possible in the future.

\section{Weak gravitational lensing of the \lya\ forest}

\subsection{The \lya\ forest as a source field}

\label{lyasource}

The IGM at redshifts $z>2$ is almost completely photoionized, by an
ultraviolet background radiation field dominated by the integrated light
of quasars (e.g., Haardt and Madau 2012). 
Simulations (Cen {\it et al.} 1994, Zhang {\it et al.} 1995, Hernquist {\it et al.} 1996).
 and analytic models (Bi 1993, Bi and Davidsen 1997) have
shown that in the standard cosmological model the forest is generated
by residual neutral hydrogen in this photoionized medium. This
 medium  fills the space between galaxies with absorbing material,
and its structure on scales larger than the Jean's scale traces
the overall matter density. The physical processes governing
the state of the IGM are simple, and its absorption properties are those
first described by Gunn and Peterson (1965), leading to its characterization as
the ``Fluctuating Gunn-Peterson Effect'' (Weinberg {\it et al.} 1997).

The \lya\ forest of absorption features was first observed in the spectra
of background quasars (Lynds and Stockton 1966).
It has also been seen in the spectra
of galaxies (e.g., Lee {\it et al.} 2014), 
but in each case provides an irregular sampling
of the IGM, consisting of one-dimensional skewers extending towards the
observer from the emitting object. The useful dimension in wavelength
from that of the  \lya\ transition to the \lyb\ transition, together with
redshift range where there are numerous quasars available means that 
redshifts $z=2-3.5$ are the most studied for cosmology. Using 
the \lya\ forest as a background field for lensing will therefore mean that
higher redshifts are available than for galaxies, the redshift distributions
of which peak between $z=0.5-1.3$ for surveys such as LSST, Euclid and WFIRST 
(Pearson {\it et al.} 2014).

The discrete nature of \lya\ forest sightlines means that a dataset made from 
a compilation of quasar or galaxy spectra will have some different statistical
properties from a true intensity map made from e.g., 21cm data. 
Density fluctuations on scales smaller than the separation between sightlines
will add a  stochastic component to the \lya\ forest signal. When three
dimensional maps from spectra are made with interpolation techniques (Pichon 
{\it et al.} 2002),
techniques such as Weiner filtering are used to deal with this noise and
yield maps which are smooth on scales below the mean sightline
spacing. The Jean's pressure smoothing scale of the forest is
around 500  comoving  $\hkpc$  at these
redshifts (Peeples {\it et al.} 2010), so that any dataset with
mean transverse sightline separation greater than this will be affected
by the stochastic component. The noise characteristics along \lya\ forest
sightlines on the other hand are dominated by photon counting noise. 
With the brightest quasars and large telescopes, spectra with per pixel
signal to noise ratios of 100 or more can be obtained (e.g.,
Kim {\it et al.} 2004). For 
large-scale surveys of quasars, however, S/N of order unity in 
pixels of size ~1 \AA\ is more typical (Lee {\it et al.} 2013). There is also a
long-wavelength component of noise which comes from continuum fitting. 
By contrast, the noise properties of 21cm observations 
are dominated by foreground removal (e.g., galactic
synchrotron), and by sky and receiver noise.  The foreground emission far overpowers 
the signal and removing it from the data will be challenging.

When dealing with  \lya\ forest clustering it is 
customary to define the ``flux overdensity'', $\df$,
where
\begin{equation}
\df=\frac{F}{\langle F \rangle} -1.
\end{equation}
$\df$ is a quantity with zero mean.

\subsection{Lensing geometry}

\begin{figure}
  \begin{center}
    \includegraphics[scale=0.33]{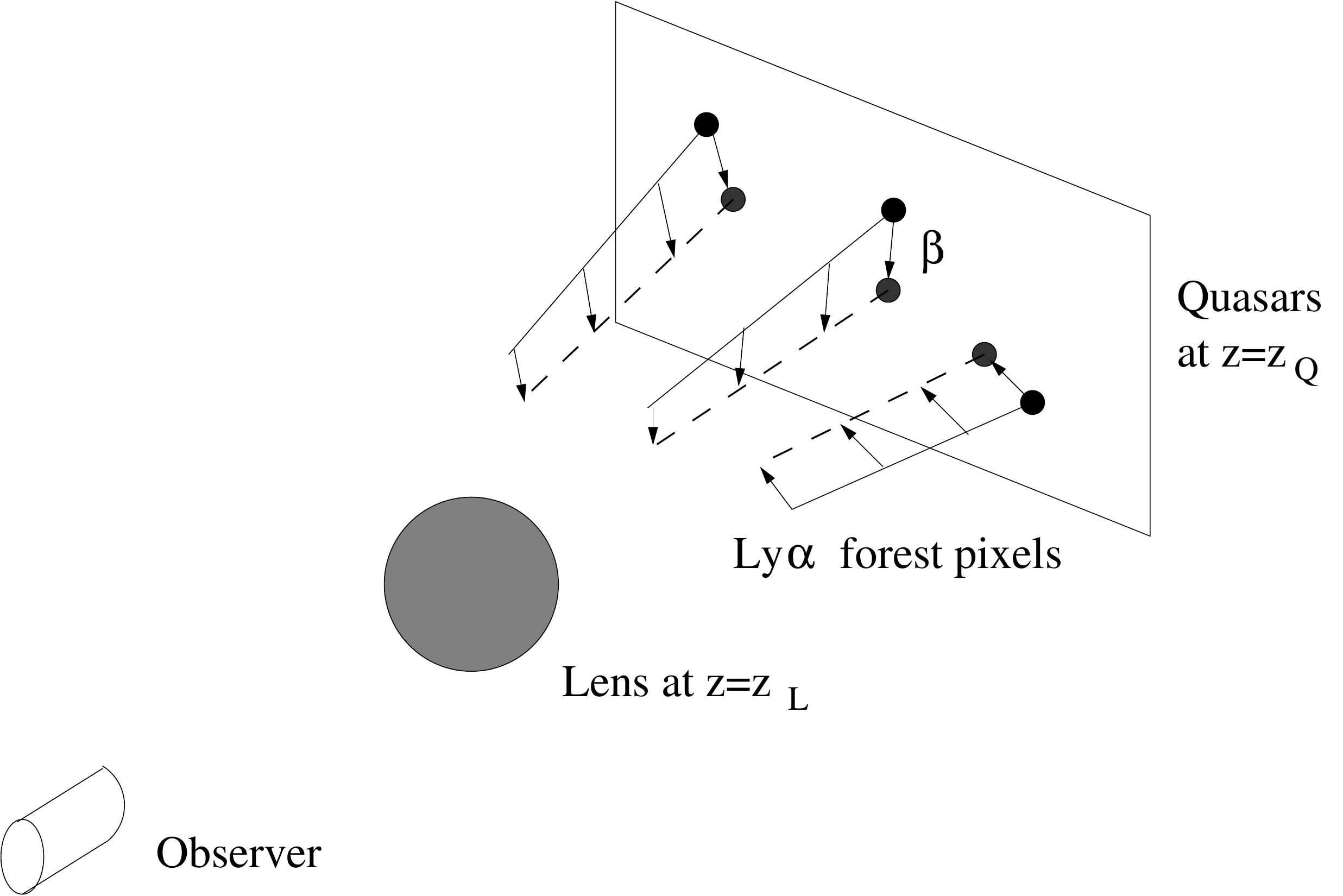}
  \end{center}
  \caption{
A cartoon illustration of the geometry of \lya\ forest lensing.
Due to lensing by the foreground object at redshift $z_{L}$, the 
angular position of a quasar at redshift $z=z_{Q}$ is 
deflected by an angle $\beta$. The associated
\lya\ forest pixels are also lensed, deflected by an angle smaller than $\beta$ (which depends on the angular size distance to the absorption in the pixel).
           }
  \label{cartoon}
\end{figure}

\begin{figure}
  \begin{center}
    \includegraphics[scale=0.33]{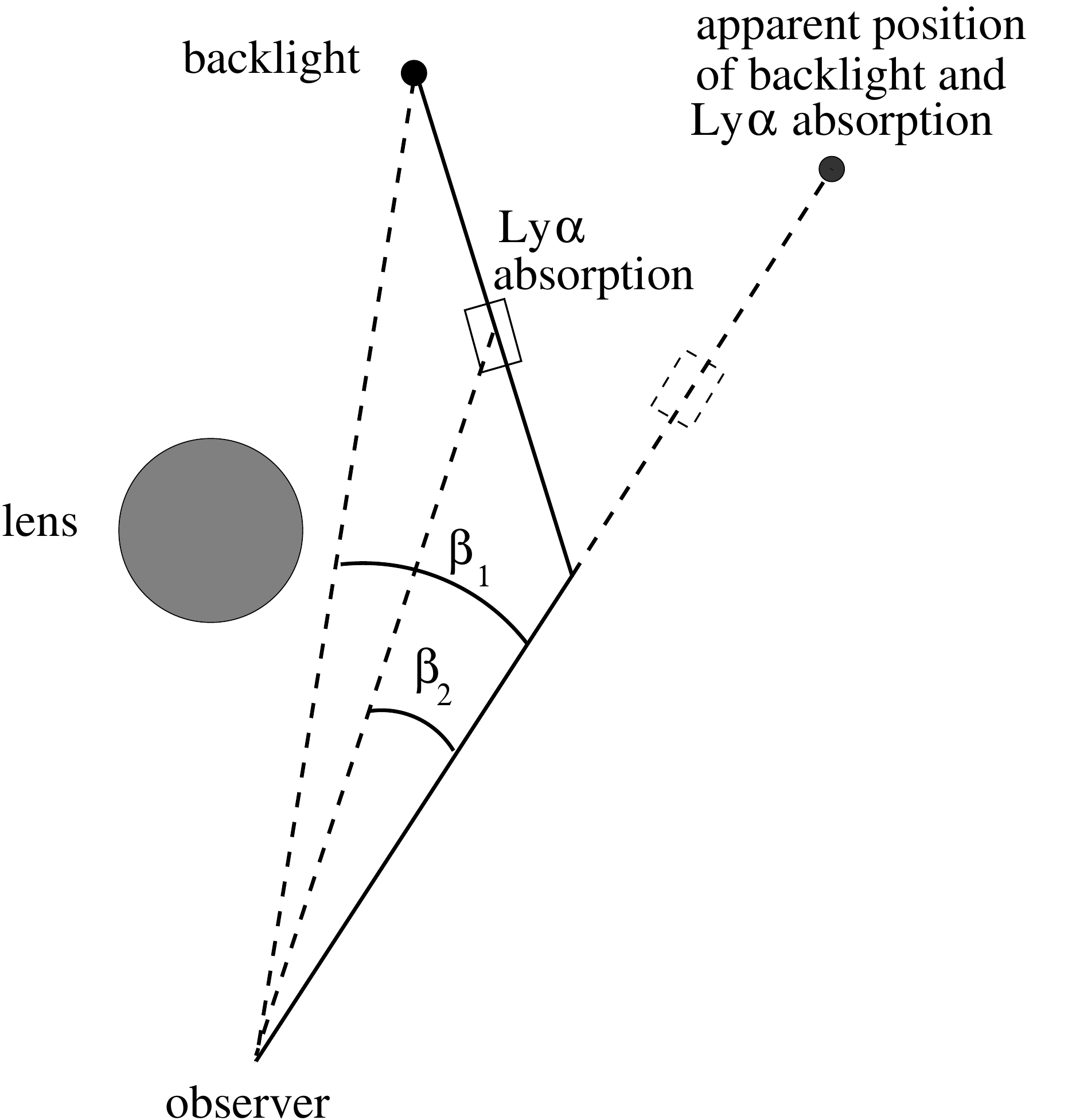}
  \end{center}
  \caption{
An illustration of lensing deflection for a single backlight (quasar or galaxy) and \lya\ forest pixel. The deflection angle for the backlight ($\beta_{1}$) is different from that for the \lya\ forest absorption ($\beta_{2}$) because they are at different redshifts (and therefore angular size distances).
           }
  \label{angdiag}
\end{figure}

The \lya\ forest  at redshift $z_{s}$ is lensed
by the matter distribution lying between us and $z_{s}$. Gravitational
lensing shifts the observed positions of points on the
sky without changing their surface brightness. 
In the case of the \lya\ forest, this means that the 
quasar or galaxy backlights move on the plane of the sky.
The angular sizes of quasars   are  magnified (through
the change in their angular sizes). This magnification of quasars through
lensing is well studied, from the first observed lenses, which created
multiple quasar images (Walsh {\it et al.} 1979), 
through weak lensing magnification of
quasars detected by cross-correlation with foreground galaxies 
(e.g., Scranton {\it et al.} 2005).
The magnification of quasars will cause some selection biases which will 
affect the topic of this paper, and which we will return to in Section
\ref{discussion}. Directly relevant to \lya\ forest lensing, however
are distortions of the angular separations between quasars, caused by 
the lensing deflection of light.  Because the quasar angular positions
move on the plane of the sky, this means that the \lya\ forest
skewers associated with each quasar also shift 
as the parent quasar does (assuming that the lensing takes place at lower redshifts 
which is a well justified approximation).  Gravitational lensing therefore distorts the "image" of the
IGM probed by the \lya\ forest without changing the transmitted flux
measured in each pixel. This is directly analogous to the effect of
lensing on 21cm emission (or the CMB), which conserves surface brightness.
In Figure \ref{cartoon} we illustrate this with a diagram, which shows the \lya\ forest
pixels being deflected in a similar fashion to the quasar backlights. In Figure \ref{angdiag} we concentrate on a single backlight and \lya\ forest pixel and show the relationship to the unlensed angular positions of both. The pixel and the backlight are at different angular size distances, and so are displaced on the sky by different angles.

If the lensing deflections are small
compared to structure in the source, the \lya\ forest 
flux overdensity $\df$  at wavelength $\lambda$ can
be expressed as a Taylor expansion of the unlensed $\df$.

\begin{equation}
\widetilde{\delta}_{F}({ \pmb {\theta}},\lambda)=
{\delta_{F}}({ \pmb {\theta}} - {\pmb { \alpha}}({\pmb {\theta}}) ,\lambda)
\simeq {\delta_{F}} ({\pmb{\theta}},\lambda) - {\pmb {\alpha}}({\pmb{\theta}})\cdot
{\pmb{\nabla_{\theta}}} \delta_{F} ({\pmb{\theta}},\lambda) + ...
\label{taylor}
\end{equation}

 This expansion is  valid in the case of the \lya\ forest, where gradients
in $\delta_{F}$ can be large, but the deflections (or deflection gradients) 
are small compared to them on all scales of interest. The deflection field
${\pmb {\beta}}( {\pmb {\theta}})$ is related to the 2D projected lensing potential via
${\pmb {\nabla \Phi}} =- {\pmb {\beta}}({\pmb {\theta}} )$,
in the weak lensing limit. This lensing potential can be
computed from the full 3D gravitational
potential (e.g, Bartelmann \& Schneider 2001) by an integration
over redshift:

\begin{equation}
\Phi=\frac{2}{c^2}\int^{z_{s}}_{0}dz\frac{D(z)D(z_{s},z)}{D(z_{s})}\phi(D(z){\pmb {\theta}}(z),z),
\end{equation}
where $D(z)$ is the angular size distance to redshift $z$ and $D(z,z')$ is the angular size distance between redshifts $z$ and $z'$.

The 3D  spectrum of \lya\ forest $\df$ fluctuations
 is given by (McDonald 2003):
\begin{equation}
P_{\delta F}(k) = b^{2} (1+\beta_{\rm RSD}\mu^2_k)^2 P_\delta(k),
\label{mcdonaldpk}
\end{equation} 
where $P_\delta(k)$ is the underlying dark matter power spectrum
and $\mu_k$ is the cosine of the angle between the
wave vector ${\rm \bf k}$ and the line of sight $\hat{z}$.  
$b$ and $\beta_{\rm RSD}$ are parameters that describe the relative bias
between flux and matter fluctuations and the
strength of redshift distortions respectively.
In the case of galaxies, which are conserved under
redshift distortions, 
$\beta_{\rm RSD} = \frac{d \ln D}{d\ln a} \simeq \Omega_m(z)^{0.55}$  where $D$ is the
linear growth rate (Kaiser 1987). For the \lya\ forest, which 
undergoes a non-linear transformation $F=e^{-\tau}$ between 
flux $F$ and optical depth $\tau$ (which is conserved),
$\beta_{\rm RSD}$ is a separate parameter. At the redshifts $z=2-3$ of interest,
$b$ is approximately $0.2$ and $\beta_{\rm RSD}$  unity (Slosar {\it et al.} 2011).

\subsection{Lensing reconstruction: Quadratic estimator}

\label{quade}

The quadratic lensing estimator for the CMB (or a slice of
 \lya\ forest pixels at single redshift)
is sensitive to variations in the power spectrum for  different 
regions of the sky.  These spatial variations result in
correlations in Fourier modes (or $C_\ell$'s) that would not
exist otherwise. 

The effect of lensing can be divided into shear that 
leads to anisotropy of the local power spectrum, and an
 isotropic magnification.
If the power-spectrum is scale free ($C_\ell \propto
\ell^{-2}$), the magnification cannot be determined
with a quadratic estimator.  Additionally, the shear cannot be measured
if the power-spectrum is constant.  The projected matter (and \lya\ forest)
power-spectrum  in the CDM model is approximately  constant on large
scales (small $\ell$) and is $C_\ell \propto \ell^{-2}$ at small scales.
A quadratic lensing  estimator will therefore
be sensitive to magnification
 on large scales, 
 shear on small scales and a combination of both for intermediate scales.

 Unlike the CMB, the \lya\ forest is observable at different
redshifts, with a range dependent on the wavelengths that 
can be observed and the availability of backlight quasars or  galaxies.
If we treat the forest as a sparsely sampled three-dimensional
dataset, we can slice it on the basis of redshift. 
  A three-dimensional lensing estimator then effectively stacks information for
 slices of the forest at different redshift.  
The lensing is coherent for different slices
 at the same angular position, but the intrinsic correlations of
the  \lya\ forest are governed by the CDM power spectrum (independent
Fourier modes in the linear theory case we are considering).
The excess coherent correlations can be attributed to lensing.
This is a very similar situation to that which occurs in 21cm lensing.

 The \lya\ forest fluctuations, $\df(\theta,z)$ , 
 in the radial direction are subject to the finite size in 
wavelength of observed pixels. They can therefore be
expressed in discrete Fourier space,
 (wave vector $k_\parallel = \frac{2\pi}{{\cal L}}j$, where ${\cal L}$
 is the depth of the observed volume).
Fluctuations perpendicular to the line of sight (wave vector
 $\mathbf{k_\perp}=\bl/{\cal D}$ where ${\cal D}$ the angular
 diameter distance to the source redshift and $\bl$ is the dual
 of the angular coordinate on the sky) are  subject to the 
limitations of the angular density of backlights. In the simplified example
cases we consider here, we will assume that there is a high enough
density of backlights that we are able to use a continuous Fourier space
representation perpendicular to the line of sight. Development of
an estimator which treats the sparseness and irregularity of angular
sampling is left to future work (Metcalf {\it et al.}, in preparation).


An optimal quadratic estimator for lensing using the CMB was developed by 
Hu and Okamoto (2002) which involves a convolution in Fourier space.   As shown by
Anderes (2013), Lewis \& Challinor (2006) and Carvalho \& Moodley
(2010), this  is equivalent to a real space product of the high-pass filtered fields. Calculating the product in real space allows
one to take advantage of Fast Fourier Transform techniques
and significantly speeds up the estimation.  
As pointed out by Lewis \& Challinor (2006) (again in the context of the CMB),  seen from this
point of view, the estimator measures the correlations in the product
of two Wiener filtered fields,  the  gradient field ${\pmb{\nabla}}{\delta}({\pmb{\theta}})$,
and a small-scale weighted  and filtered version of $\delta({\pmb{\theta}})$.

In the context of \lya, or 21cm emission, this estimator for the gravitational potential can be written as (dropping the $F$ subscript in $\delta_F$ for simplicity)
\begin{align}
 \hat{\Phi}(\bL) & = -\frac{N^\phi_{\bL} }{\Omega_s} (i\bL)\cdot \sum_{k_p} \left[ \sum_{\pmb\theta} e^{-i\bL\cdot \pmb{\theta} } F_{\pmb\theta} \nabla G_{\pmb \theta} \right]_{k_p} 
 \label{eq:estimator}
\end{align}
where $F_{\pmb\theta}$ and $G_{\pmb \theta}$ are the DFT of
\begin{align}
F_{\bl,k_p} = \frac{\delta_{\bl,k_p}}{C^{\rm tot}_{\ell,k_p}} ~~~,~~~ G_{\bl,k_p} = \frac{C_{\ell,k_p}\delta_{\bl,k_p}}{C^{\rm tot}_{\ell,k_p}}.
\end{align}
This is a configuration space version of the estimator given by ZZ06 as derived in 
Romeo {\it et al.}, (2017).  The estimated noise in this estimator is
 \begin{equation}
N^\phi_{\bL} = \left[ \frac{1}{2\Omega_s} \sum_{\bl,j}^{j_{\rm max} } \frac{[\bell \cdot \bL C_{\ell,j}+\bL \cdot
    (\bL-\bl) C_{|\ell-L|,j}]^2}{ C^{\rm tot}_{\ell,j}C^{\rm
      tot}_{|\bl-\bL|,j}}\right]^{-1}, 
\end{equation}
 where 
\begin{equation}
\label{eq:Cellj}
C_{\ell,j} = \frac{P_{\delta_{f}}(\sqrt{(\ell/{\cal D})^2+
(j2\pi/{\cal L})^2})}{{\cal D}^2 {\cal L}} =  [\bar{T}(z)]^2  P_{\ell,j}.
\end{equation}
$C^{\rm tot}_{\ell,j}$ is the total power spectrum:  $C_{\ell,j}$, plus the power spectrum of the noise, $C^N_{\ell,j}$.  For further details we refer the reader
to Romeo (2015) and Romeo et al . (2017) for details related to this form of the estimator.

In the present paper, as we make the simplifying assumption that the \lya\ sightlines are 
distributed on a grid,  the 21cm reconstruction techniques used in Romeo (2015) are
directly applicable. 
We will also simulate a higher density of sightlines than most current surveys. 

\section{Test on gridded data.}
\label{testg}

In this section, we carry out a test of reconstruction of the lensing 
mass distribution in an  idealized and simplified case. 
These simplifications include a high density of 
parallel sightlines,  with all backlights (quasar or galaxy)
on a grid, and at the same redshift. We also use Gaussian random
fields to model both the foreground lens and the source fields.
Our aim is to illustrate the potential of \lya\ forest lensing. We leave more realistic
simulations and the development of mass estimators which are able to deal
with  more complex geometries (and sparser sightlines)  to future work.
The generation of the source and lens fields are modeled largely on
techniques used to simulate 21cm fluctuations in 
Romeo (2015) and Romeo {\it et al.} (2017), and the reader is referred to
them for more details.

\subsection{Source field}

We use the \lya\ forest at redshifts between $z=2$ and $z=3$ as a 
source field. The forest is generated by mass fluctuations which are within
an order of magnitude of the cosmic mean (Bi, 1993), and we choose
to model the field using linear theory, as a Gaussian random field.
In the future, lognormally transformed fields (Bi \& Davidsen
1997, Le Goff, 2011), large hydrodynamic 
simulations (e.g., Cisewski et al 2014), or hybrid
methods (Peirani {\it et al.} 2014)  
should be used to model the truly quasilinear nature
of the forest (and to test the response of the estimators).

The \lya\ forest flux fluctuations
$\df$ are biased with respect to the matter fluctuations
$\df$. The power spectra of the two are related in linear theory
as in equation \ref{mcdonaldpk}.  Gaussian random density
fields into \lya\ forest flux fields. This direct conversion of
an isotropic field into another is a simplification, as we
are not including only Kaiser-form redshift distortions.  This neglects the
effects of velocity dispersion and thermal broadening, both of which act to smooth structure
along the line of sight. Again, these are not expected
to qualitatively affect our test and are left to future work.

The linear power spectrum we use is that of a Cold Dark Matter universe with
the parameters given by the Planck 2015 data release (Ade {\it et al.}, 2016).
We simulate a sky area of 15 by 15 degrees, with $468^{2}$
angular pixels. The pixel size of 1.8 arcmins corresponds to a
comoving length scale of 2.2 $\hmpc$ comoving at redshift $z=2.5$. This is 
larger than the relevant Jeans smoothing scale of the forest 
(of order 0.5 $\hmpc$ at this redshift, Peeples {\it et al.} 2010). 
The effects of fluctuations on scales intermediate between our 
resolution and the Jeans scale are therefore missing from our simulation.
 In order to include their effects, we could simulate the forest field with a high enough
density of grid points to resolve the Jeans scale and then subsample
sightlines to yield the required sightline density.  An alternative, 
would be to assume that the missing fluctuations would add to the grid in a
similar fashion to noise.  In practice, neither is necessary for the illustrative example in this
paper because we average the forest pixel in slices of thickness greater than $10 \hmpc$ comoving, and this averaging smooths out  the small scale fluctuations.  The rms fluctuations
on $\sim 10 \hmpc$ scales are approximately $\sigma  = 0.04$,  based on the  correlation function  measurements  of Croft {\it et al.} ( 2002) at $z=2.5$.,  and this is comparable to the rms fluctuations in our generated field.

\subsection{Lens field}
\label{sec:lensfiled}

\begin{figure}
  \begin{center}
    \includegraphics[scale=0.35]{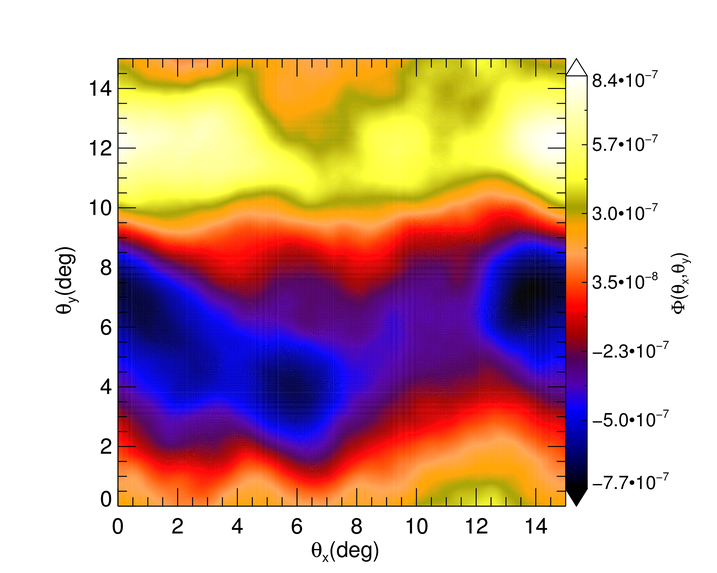}
  \end{center}
  \caption{ The input lens potential field for the
reconstruction test. The sky area covered is 15 by 15 degrees
and the lens is at redshift $z=1$.
           }
  \label{inputlens}
\end{figure}

As we will see, the lensing field will be detected only on large angular scales 
where the matter structure responsible for the lensing is linear to a very 
good approximation and thus well represented by a Gaussian random field.   
We create a realization of the lensing potential by generating random Gaussian 
distributed Fourier modes with the power spectrum expected in the same cosmology 
used to simulated the lya\ forest source fields.  An image of the lensing potential can be seen in figure~\ref{inputlens} for a 15 by 15 degree patch of the sky. The deflection is found by taking the 
gradient of the potential use the FFT (Fast Fourier Transform).  Points on the source planes 
are then displaced by the deflection and interpolated back onto a regular grid to 
get the lensed image.  A plane approximation is used here which is not overly accurate on the scales considered here, but for our purposes a less than 10\% inaccuracy is not important.

\begin{figure}
  \begin{center}
    \includegraphics[scale=0.35]{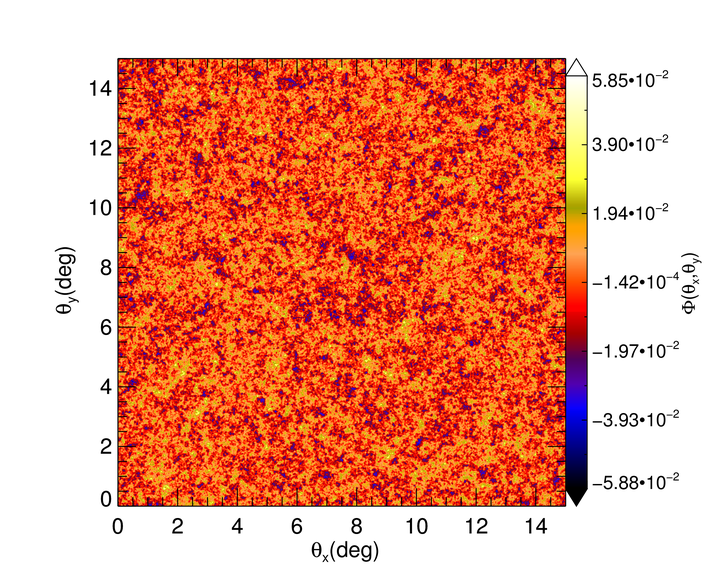}
  \end{center}
  \caption{A slice through the unlensed \lya\ forest source field at 
redshift $z=2.5$. The \lya\ forest in our mass reconstruction test
covers redshifts $z=2-3$ and was generated on a grid, representing an 
idealized grid of quasar spectra. The color scale represents the 
flux overdensity, $\df$.
           }
  \label{unlensedsource}
\end{figure}

\begin{figure}
  \begin{center}
    \includegraphics[scale=0.35]{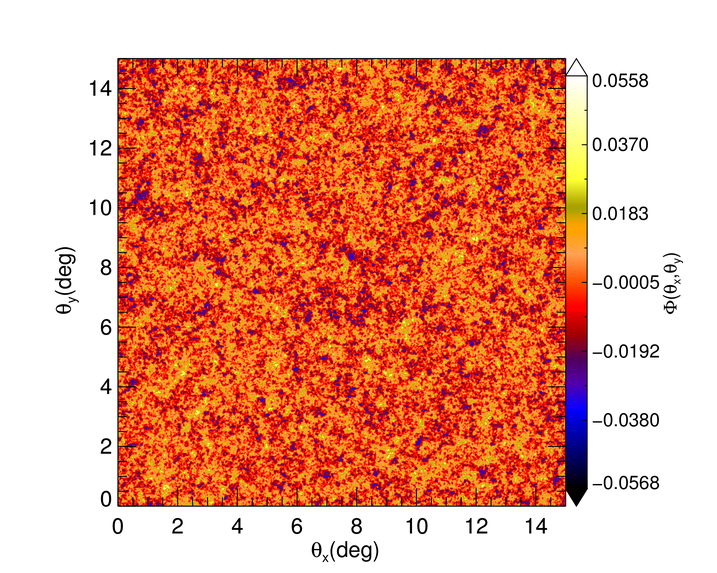}
  \end{center}
  \caption{A slice (at redshift $z=2.5$, the 
same as in Figure \ref{unlensedsource}) 
through the lensed \lya\ forest source field.
in our mass reconstruction test. 
           }
  \label{lensedsource}
\end{figure}

\begin{figure}
  \begin{center}
    \includegraphics[scale=0.35]{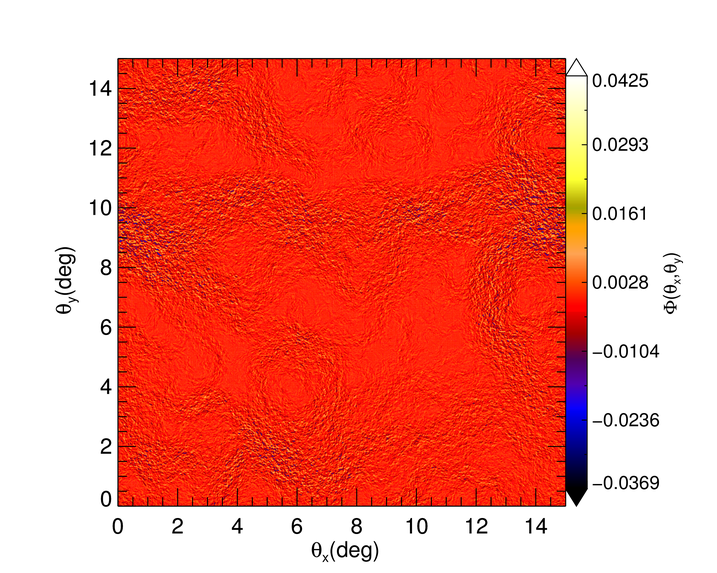}
  \end{center}
  \caption{The difference between the lensed and unlensed source fields
shown in Figures \ref{unlensedsource} and \ref{lensedsource}.
           }
  \label{diff}
\end{figure}

\subsection{Mass reconstruction}

The lensing potential estimator, (\ref{eq:estimator}), is applied to the lensed \lya\ data cube.  We approximate the noise in each spectral pixel as being independent with a power spectrum
\begin{equation}
C^N_{\ell,j} = \frac{\sigma^2}{{\cal D}^2 {\cal L}} \frac{V_\alpha}{N_\alpha}
\end{equation}
where $V_\alpha$ is the total volume spanned by the \lya\ data and $N_\alpha$ is the total number of pixels.
This ignores variations in the noise between different backlights and at different redshifts.  It also 
ignores possible correlations between pixels coming from the fitting and subtraction of the backlights' continuum spectra.  These complications are unlikely to present a significant problem, having already been  tackled by groups measuring baryon oscillations from the forest. We will  addressed them in the context of  forest lensing in future work.

\subsection{Results}
\label{results}

\begin{figure}
  \begin{center}
    \includegraphics[scale=0.35]{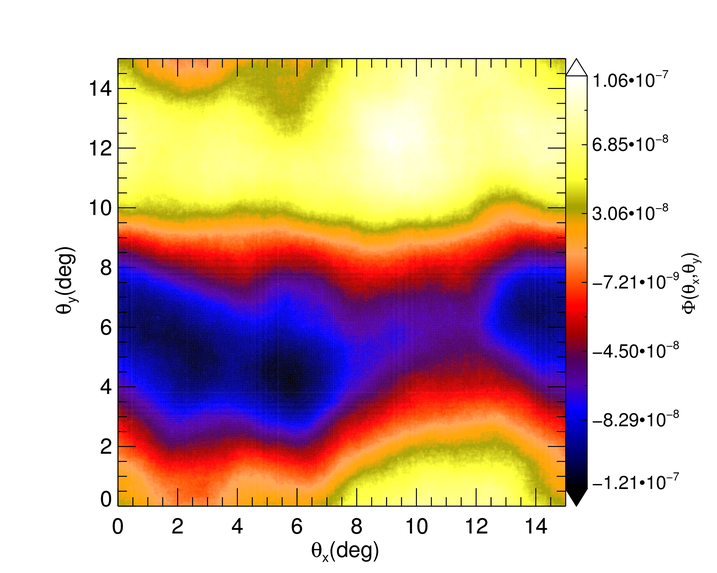}
  \end{center}
  \caption{The recovered lens potential field from our test of
matter reconstruction from forest. The field should be compared with the
input lens potential from Figure \ref{inputlens}           }
  \label{recovered lens}
\end{figure}

We preformed simulations to see how well an image of the lensing potential could be 
recovered.  We used a 15 by 15 degree field with 512 pixels on a side.  The input lensing potential is shown in figure~\ref{inputlens} as described in section~\ref{sec:lensfiled}.  A single slice through the \lya\ is shown in figure~\ref{unlensedsource}.   Figure~\ref{lensedsource} shows the same slice after being lensed.  The difference is difficult to see by eye so Figure~\ref{diff} shows the difference between these maps.  One can see clear patterns that are related to the input potential.  Figure~\ref{recovered lens} shows the recovered lensing potential.  By comparing \ref{recovered lens} with \ref{inputlens}, it can be seen that the large scale morphology of the potential fluctuations on a scale of several degrees are well recovered.

\section{Summary and discussion}

\label{sumconc}

\subsection{Summary}
We have introduced the concept of weak gravitational lensing of the \lya\ 
forest as a potentially detectable effect. 
Using simulations of Gaussian random fields as both source and lens,
and simplified planar geometry,
we have shown that it will likely be possible
to reconstruct images of the foreground mass distribution given
a sufficiently large and dense sample of quasar and galaxy spectral data.

\subsection{Discussion}

\label{discussion}

Weak lensing of the \lya\  forest has some potential
advantages over other lensing probes. Compared to galaxies, the forest offers
a higher redshift source field, and well known source redshifts. Compared
to the CMB, the data is three dimensional, with many source planes, and
compared to 21cm, there is data available now, and the signal is
not dominated by astrophysical foregrounds. Unfortunately the forest has
several disadvantages, such as the fact that it sparsely samples the 
source density field. Without further tests with more realistic geometries
and samples it is not clear how useful the \lya\ forest will be as a
cosmological tool. Neverthless, relevant data will be taken in the future, and 
in the best scenario the lensing of the forest could enable robust 
cosmological constraints  on quantities such as the neutrino mass, as
has been forecast for CMB lensing (e.g., Abazajian {\it et al.} 2016).

\begin{table*}
\begin{tabular}{|l|l|l|l|l|}
\hline
Dataset   & When      & Area            & N_{\rm spectra} & mean separation \\ \hline
BOSS DR12 & 2016      & 10,000 sq. deg. & 160,000            & 15 arcmin       \\
eBOSS     & 2014-2018 & 7,500 sq. deg.  & 270,000            & 10 arcmin       \\
CLAMATO   & 2014-2018 & 0.8 sq. deg.    & 1,000              & 1.7 arcmin      \\
WEAVE     & 2018-2020 & 6,000 sq. deg.  & 400,000            & 7.5 arcmin      \\
DESI      & 2018-2023 & 14,000 sq. deg. & 770,000            & 8.1 arcmin      \\
Subaru PFS & 2019-2022  & 15 sq. deg. &  7,400   & 2.7 arcmin \\ 
MSE       & 2025-     & 1,000 sq. deg.  & 1,000,000          & 1.9 arcmin     \\\hline
\end{tabular}
\centering
\caption{
Some relevant parameters for future \lya\ forest observational
datasets. Of these, BOSS (Dawson {\it et al.} 2013) has been completed,
eBOSS (Dawson {\it et al.} 2016) and CLAMATO (Lee {\it et al.} 2014, using
galaxy spectra) are ongoing, 
and WEAVE (Dalton {\it et al.} 2012) and DESI (Aghamousa {\it et al.} 2016)
are about to start.  A Subaru PFS (Takada {\it et al.} 2016) survey may be carried out
for IGM tomography and some possible parameters are given. The survey labelled MSE
is a potential star forming galaxy with the proposed
Mauna Kea Spectrosopic Explorer 
instrument$^{0}$.
\label{obs}
}
\end{table*}

\footnotetext{http://mse.cfht.hawaii.edu/}

So far we have not discussed systematic biases in mass reconstruction from
forest lensing. One can imagine that the magnification of quasar and galaxy
backlights by the same lensing mass one is trying to reconstruct could cause
biases. Quasar lensing magnification has been the subject of extensive
studies (e.g., Scranton {\it et al.} 2005), and how lensing will
affect the locally determined luminosity function is well understood.
The selection of sightlines to quasars or galaxies will depend on their
magnification, and as this depends on the foreground mass, the sampling
and the signal to noise of source spectra will be correlated with the 
lensing mass.
The effect of this on measurement bias on mass reconstruction 
from the forest should be be
explored (see the related effect on galaxy lensing studied by
 Liu {\it et al.} 2014)
 The same issues will also arise because of lenses in the forest 
itself. As the lensing kernel is broad, the \lya\ forest could 
have a detectable influence on the observed magnitude of quasars. Loverde
{\it et al.} (2010) have studied how this causes a measurement bias and also how
it could be detected by cross-correlating \lya\ forest flux statistics
and quasar magnitudes.

Many of the problems which affect measurement of cosmic shear with galaxy
shapes are unlikely to affect the \lya\ forest.  For example, the 
sphericalisation and bias due to atmospheric seeing (e.g., Weinberg
{\it et al.} 2013) is likely to be 
acting on scales much smaller than the separation between 
galaxy and quasar sightlines.
Some non-lensing alignments are likely to persist, however, such as
gravitational-intrinsic alignment correlations (Hirata and Seljak 2004).

Many improvements must be made to the simplified simulations we have used 
here in order to truly test \lya\ forest lensing, and also to investigate
the strength of the biases we have mentioned above. These improvements 
should cover both the source and the lens fields. In the case of the
\lya\ forest source, we have used Gaussian random fields, but the 
 observed forest, being in the quasi-linear regime
 has a close to log normal probability distribution (Bi and Davidsen 1997). 
Log normal transformed Gaussian random fields  could be  
used to make more realistic simulations
as well as physical effects such as thermal broadening. Simulation
techniques exist (e.g., Peirani et al 2014) which can be used to make
extremely large forest data sets by combining dark matter n-body simulations
and information from hydrodynamic models. In the future, fully hydrodynamic
simulations will eventually be run which cover the large sky area
needed.

We have also used Gaussian random lens fields in this paper. Models
based on raytraced n-body simulations are a more realistic alternative.
The Multi-Dark Lens simulations (Giocoli {\it et al.} 2016) are available, 
for example, and the GLAMER (Metcalf and Petkcova
2014) raytracing code can be used to 
compute lens maps at higher redshifts than have been done for galaxy
sources.

Mock \lya\ forest observations are also needed for true tests, and these
should include realistic geometries, with sightlines converging on 
an observer, and with non-uniform sampling on the plane of the sky.
The number density of sightlines should also be varied widely, as the
density used in this work (of the order of 1 sightline per square arcminute)
is at the extreme high end, two orders of magnitude larger than 
currently available datasets such as BOSS and eBOSS (Dawon et al 
2013, 2016). We have also
only simulated a few square degrees of sky area, and can therefore hope that
statistical studies of larger areas will benefit from the greater total
number of spectra. This again is something which larger, more realistic
mocks will help determine. Mock observations should also include various
sources of noise, from photon shot noise to large-scale variations
caused by continuum fitting errors. Metal lines should
also be evaluated as a source of contamination as they are in
\lya\ forest clustering measurements (Bautista et al 2016).

In order to deal with realistic geometries, a more inclusive estimator
is needed. We are currently developing such an estimator, and plan
to apply it to observational data.
The field of \lya\ forest observations is opening up rapidly as
large datasets are being made publicly available and new instruments
are being planned. The largest current dataset in terms of number of
spectra forms part of the twelfth data release (DR12) of SDSS/BOSS.
We show in Table \ref{obs} some current and future \lya\ forest datasets
which could be used to carry out lensing studies. Although the mean separation
of sightlines in current and ongoing surveys is relatively large ($\sim
10$ arcmins or more), values close to one arcminute could be achievable with
highly multiplexed observations of galaxy spectra on future large
telescopes. Such observations may offer a route to precision cosmology
from \lya\ forest lensing.

\section*{Acknowledgments}
 
RACC is supported by NASA ATP grant NNX17AK56G.
AR and RBM have been supported partly through project GLENCO, funded under the European Seventh Framework Programme, Ideas, Grant Agreement n. 259349.
{}

\end{document}